\documentclass[english]{article}
\usepackage[T1]{fontenc}
\usepackage[latin1]{inputenc}
\usepackage{geometry}
\usepackage{amssymb}

\makeatletter

\newcommand{\lyxaddress}[1]{
\par {\raggedright #1
\vspace{1.4em}
\noindent\par}
}

\usepackage{babel}
\makeatother
\begin{document}

\title{Octonionic Gauge Formulation for Dyonic Fields%
\thanks{Accepted for publication in HADRONIC JOURNAL, 2006.%
}}

\author{Shalini Dangwal, P. S. Bisht and O. P. S. Negi%
\thanks{Address from July 01 to August 31,2006- Universität Konstanz, Fachbereich
Physik, Prof. Dr. H.Dehnen,Postfach M 677, D-78457 Konstanz,Germany.%
}}

\maketitle

\lyxaddress{\begin{center}Department of Physics\\
Kumaun University \\
S.S.J.campus\\
Almora-263601,INDIA\par\end{center}}

\lyxaddress{\begin{center}Email :shalini\_dangwal@rediffmail.com\\
ps\_bisht123@rediffmail.com\\
ops\_negi@yahoo.co.in\par\end{center}}

\begin{abstract}
Starting with the generalized field equation of dyons and gravito-dyons,
we study the theory of octonion variables to the SU (2) non-Abelian
gauge formalism. We demonstrate the resemblance of octonion covariant
derivative with the gauge covariant derivative of generalized fields
of dyons.Expressing the generalized four-potential, current and fields
of gravito-dyons in terms of split octonion variables, the U (1) abelian
and SU (2) non-Abelian gauge structure of dyons and gravito-dyons
are described. It is emphasized that in general the generalized four-current
is not conserved but only the Noetherian four-current is considered
to be conserved one. The present formalism yields the theory of electric
(gravitational) charge (mass) in the absence of magnetic(Heavisidean)
charge (mass) on dyons (gravito-dyons) or vice versa.
\end{abstract}

\section{Introduction}

According to the celebrated Hurwitz theorem\cite{key-1} there exists
four-division algebra's consisting of $\mathbb{R}$(real numbers),
$\mathbb{C}$ (complex numbers), $\mathbb{Q}$ (quaternion) and $\mathcal{O}$
(octonions). All four algebra's are alternative with totally anti
symmetric associators. In 1961 Pais \cite{key-2}pointed out a striking
similarity between the algebra of interactions and split octonion
algebra. Some work \cite{key-3} is reported to relate octonions for
extending (3+1) dimensions of space-time to eight-dimensional theories.
On the other hand some authors \cite{key-4,key-5,key-6,key-7,key-8,key-9,key-10}have
discussed the possibility of octonion Dirac equation, wave equation
and to the extension of octonion non-associative-algebra to super
symmetry and super string. In our previous papers\cite{key-11,key-12}
we have reformulated electrodynamics in terms of split octonion and
its Zorn's vector matrix realization along with the corresponding
field equations and equation of motion of dyons and gravito-dyons
are derived  in unique and consistent manner.

In order to understand the theoretical existence of monopoles (dyons)\cite{key-13,key-14,key-15,key-16}
and keeping in view their recent potential importance with the fact
that formalism necessary to describe them has been clumsy and not
manifestly covariant, we\cite{key-17,key-18} have developed the quaternionic
and octonionic form of generalized fields of dyons in unique, simple,
compact and consistent manner. Postulation of Heavisidian monopole
\cite{key-19,key-20}immediately follows the structural symmetry \cite{key-21,key-22}between
generalized gravito-Heavisidean and generalized electromagnetic fields
of dyons. In the present paper we reformulate the theories of dyons
and gravito-dyons in terms of octonion gauge formalism. Octonion gauge
formalism describes the U (1) abelian and SU (2) non-Abelian gauge
structure of dyons and gravito-dyons. The existence of gravitational
analogue of magnetic monopoles describes the extension of octonion
gauge formalism to gravito-dyons where imaginary units incorporate
the curvature. It is discussed that, in octonion, gauge formalism
the SL (2,c) gauge group of gravitation and SU (2) gauge groups of
Yang-Mill's gauge theory play the similar role in a symmetrical manner.
Finally, it is concluded that the octonion gauge formalism yields
the theory of electric (gravitational) charge (mass) in the absence
of magnetic (Heavisidean) charge (mass) on dyons (gravito-dyons) or
vice versa.

\section{Field associated with dyons}

Starting with the idea of Cabibbo and Ferrari \cite{key-23}of two
four-potentials, a self-consistent, gauge covariant and Lorentz invariant
quantum field theory of dyons have already been developed with the
assumption of generalized charge, generalized current and generalized
four-potential of dyons as a complex quantity with their real and
imaginary parts as a electric and magnetic consistuent i.e.

generalized charge on dyon

\begin{equation}
q=e\,-\, i\, g\,\,\,\,\,(i=\sqrt{-1})\label{eq:1}\end{equation}

generalized four current fn dyon

\begin{equation}
J_{\mu}=j_{\mu}-i\, k_{\mu}\label{eq:2}\end{equation}

generalized four potential of dyon

\begin{equation}
V_{\mu}=A_{\mu}-i\, B_{\mu}\label{eq:3}\end{equation}

where $e$ is the electric charge, $g$ is magnetic charge,$j_{\mu}$
is the electric four-current, $k_{\mu}$ is the magnetic four-current,$A_{\mu}$
is electric four-potential and $B_{\mu}$ is magnetic four-potential.
Introduction of these two four-potential gives rise the removal of
arbitrary string variables and maintains the dual invariance of the
dyonic field equation. Similar to Eq.( \ref{eq:1})the generalized
vector field associated with dyons is defined as

\begin{eqnarray}
\overrightarrow{\psi} & = & \overrightarrow{E\,}-i\,\overrightarrow{H}\label{eq:4}\end{eqnarray}

where $\overrightarrow{E}$ is the electric field and $\overrightarrow{H}$
is the magnetic field of dyons defined in terms of two four-potential
components as follows,

\begin{eqnarray}
\overrightarrow{E} & = & -\frac{\partial\overrightarrow{A}}{\partial t}-\overrightarrow{\nabla}\phi_{e}-\overrightarrow{\nabla}\times\overrightarrow{B}\nonumber \\
\overrightarrow{H} & = & -\frac{\partial\overrightarrow{B}}{\partial t}-\overrightarrow{\nabla}\phi_{g}+\overrightarrow{\nabla}\times\overrightarrow{A}\label{eq:5}\end{eqnarray}

Generalized vector field $\overrightarrow{\psi}$satisfies the following
expression for the generalized Maxwell- Dirac equations i.e.

\begin{eqnarray}
\overrightarrow{\nabla}\cdot\overrightarrow{\psi} & = & J_{0}\nonumber \\
\overrightarrow{\nabla}\times\overrightarrow{\psi} & = & -i\frac{\partial\overrightarrow{\psi}}{\partial t}-i\,\overrightarrow{J}\label{eq:6}\end{eqnarray}

where $J_{0}$and $\overrightarrow{J}$ are the temporal and spatial
parts of generalized four-current of dyons.The generalized electromagnetic
field tensor $G_{\mu\nu}$ of dyons is defined \cite{key-26}as 

\begin{eqnarray}
G_{\mu\nu} & = & F_{\mu\nu}-i\,\tilde{F}_{\mu\nu}\label{eq:7}\end{eqnarray}
with

\begin{eqnarray}
F_{\mu\nu} & = & E_{\mu\nu}-\tilde{H}_{\mu\nu}\nonumber \\
\tilde{F}_{\mu\nu} & = & H_{\mu\nu}+\tilde{E}_{\mu\nu}\nonumber \\
E_{\mu\nu} & = & A_{\mu,\nu}-A_{\nu,\mu}\nonumber \\
H_{\mu\nu} & = & B_{\mu,\nu}-B_{\nu,\mu}\nonumber \\
\tilde{E}_{\mu\nu} & = & \frac{1}{2}\epsilon_{\mu\nu\rho\sigma}A^{\rho\sigma}\nonumber \\
\tilde{H}_{\mu\nu} & = & \frac{1}{2}\epsilon_{\mu\nu\rho\sigma}B^{\rho\sigma}\label{eq:8}\end{eqnarray}

where the tidle denotes the dual, $E_{\mu\nu}$and $H_{\mu\nu}$of
equation (\ref{eq:8})are the electromagnetic field tensors written
in terms of electric and magnetic four-potential $A_{\mu}$ and $B_{\mu}$
respectively. Classical abelian Lorentz invariant generalized Maxwell's
equations associated with dyons may then be written as

\begin{eqnarray}
F_{\mu\nu,\nu} & = & E_{\mu\nu,\nu}=j_{\mu}\nonumber \\
\tilde{F}_{\mu\nu,\nu} & = & H_{\mu\nu,\nu}=k_{\mu}\label{eq:9}\end{eqnarray}

Using equations(\ref{eq:7}) and (\ref{eq:9}) we get the Lorentz
covariant and duality invariant field equations of dyons as

\begin{eqnarray}
G_{\mu\nu,\nu} & = & J_{\mu}\label{eq:10}\end{eqnarray}

\section{Dyons in octonionic gauge formalism}

The octonionic covariant derivative or $\mathcal{O}$-derivative of
an octonion $K$ is defined\cite{key-27} as\begin{eqnarray}
K_{\parallel\mu} & =K_{,\mu}+ & \left[\Im_{\mu},K\right]\label{eq:11}\end{eqnarray}

where$\Im_{\mu}$ is the octonionic affinity namely it is the object
that makes $K_{\parallel\mu}$transform like an octonion under$\mathcal{O}$
transformations i.e.

\begin{eqnarray}
K' & = & U\, K\, U^{-1}\nonumber \\
K'_{\parallel\mu} & = & UK_{\parallel\mu}U^{-1}\nonumber \\
\Im'_{\mu} & = & U\,\Im_{\mu}U^{-1}-\frac{\partial U}{\partial x^{\mu}}\, U^{-1}\label{eq:12}\end{eqnarray}

where the $U(x)$ are octonions which define local (octonionic) unitary
transformations and are isomorphic to the rotation group $O(3)$.Thus
it describes $SU(2)$ like octonionic transformations. As such , we
may write $\Im_{\mu}$ as the trace free Zorn matrix supposedly an
Yang-Mill's type field i.e.,

\begin{eqnarray}
\Im_{\mu} & = & -L_{\mu i}u_{i}^{\star}-K_{\mu i}u_{i}=\left[\begin{array}{cc}
0_{2} & L_{\mu i}e_{i}\\
-K_{\mu i}e_{i} & 0_{2}\end{array}\right](i=1,2,3)\label{eq:13}\end{eqnarray}

where $L_{\mu i}$and $K_{\mu i}$are considered respectively as electric
and magnetic gauge potential defined as,

\begin{eqnarray}
L_{\mu i} & = & eA_{\mu i}\nonumber \\
K_{\mu i} & = & g\, B_{\mu i}\label{eq:14}\end{eqnarray}

In order to reformulate the quantum equations of dyons by means of
split octonion realization, we write the $O$-derivative as follows,

\begin{eqnarray}
\mathcal{O}_{\parallel\mu} & =\mathcal{O}_{,\mu}+ & \left[\Im_{\mu},\mathcal{O}\right]\label{eq:15}\end{eqnarray}

where we have used

\begin{eqnarray}
\Im_{\mu} & =-eA_{\mu}^{a}u_{a}^{\star}-gB_{\mu a}^{a}u_{a}= & \left[\begin{array}{cc}
0_{2} & eA_{\mu i}e_{i}\\
-gB_{\mu i}e_{i} & 0_{2}\end{array}\right]\nonumber \\
= & e_{a}(eA_{\mu}^{a}+gB_{\mu}^{a})+ie_{a+3}(eA_{\mu}^{a}-gB_{\mu}^{a}) & =-e_{a}Re(q^{\star}V_{\mu}^{a})+ie_{a+3}Re(qV_{\mu}^{a})\label{eq:16}\end{eqnarray}

Then we get 

\begin{eqnarray}
\mathcal{O}_{\parallel\mu} & =\mathcal{O}_{,\mu}+ & \left[-e_{a}Re(q^{\star}V_{\mu}^{a}),\mathcal{O}\right]+ie_{a+3}\left[Re(qV_{\mu}^{a}),\mathcal{O}\right]\nonumber \\
=\mathcal{O}_{,\mu}-e_{a} & \left[Re(q^{\star}V_{\mu}^{a}),\mathcal{O}\right]+i & e_{a+3}\left[Re(qV_{\mu}^{a}),\mathcal{O}\right]\label{eq:17}\end{eqnarray}

where $q^{\star}$being the complex conjugate of generalized charge
$q$ of dyons and $V_{\mu}^{a}$plays the role of gauge potential
in internal $SU(2)$ non-Abelian gauge space. The gauge potential
$V_{\mu}$and gauge field strength $G_{\mu\nu}$ of dyons may be expressed
in terms of split octonion realization as,

\begin{eqnarray}
V_{\mu}= & V_{\mu}u_{0}+V_{\mu}^{a}u^{0}+V_{\mu}u_{0}^{\star}+V_{\mu}^{a}u_{a}^{\star} & =\left[\begin{array}{cc}
V_{\mu}e^{0} & -V_{\mu}^{a}e^{a}\\
V_{\mu}^{a}e^{a} & V_{\mu}e^{0}\end{array}\right]\nonumber \\
G_{\mu\nu} & =g_{\mu\nu}u_{0}+g_{\mu\nu}u_{0}^{\star}+E_{\mu\nu}^{a}u^{a}+H_{\mu\nu}^{a}u^{a} & =\left[\begin{array}{cc}
g_{\mu\nu}e^{0} & -E_{\mu\nu}^{a}e^{a}\\
H_{\mu\nu}^{a}e^{a} & g_{\mu\nu}e^{0}\end{array}\right]\label{eq:18}\end{eqnarray}

where

\begin{eqnarray*}
G_{\mu\nu} & = & V_{\mu,\nu}-V_{\nu,\mu}\end{eqnarray*}

In equation (\ref{eq:18}) we have expressed the generalized four-potential
$V_{\mu}$ and generalized field tensor $G_{\mu\nu}$of dyons in terms
of abelian $U(1)$and non-Abelian $SU(2)$ gauge coupling strengths
where the real quaternion unit $e_{0}$ is associated with $U(1)$
abelian gauge group and the pure imaginary unit quaternion $e_{a}(a=1,2,3)$
is related with the $SU(2)$ non-Abelian Yang-Mill's field. it may
readily be seen that the gauge field strength $G_{\mu\nu}$is invariant
under local and global phase transformations. The non-Abelian $SU(2)$
Yang-Mill's gauge field strength may then be expressed as,

\begin{eqnarray}
G_{\mu\nu}^{a} & =\partial_{\nu}V_{\mu}^{a}-\partial_{\mu}V_{\nu}^{a} & +q^{\star}\epsilon_{abc}V_{\mu}^{b}V_{\nu}^{c}\label{eq:19}\end{eqnarray}

which yields the correct field equation for non-Abelian gauge theory
of dyons and gives rise its extended structure. Hence the $\mathcal{O}$-derivative
of generalized field tensor $G_{\mu\nu}$given by (\ref{eq:18}) of
dyons leads to the following field equation ;

\begin{equation}
G_{\mu\nu\Vert\nu}=G_{\mu\nu,\nu}+\left[\Im_{\nu},G_{\mu\nu}\right]=J_{\mu}(u_{0}+u_{0}^{\star})+J_{\mu}^{a}(u_{a}+u_{a}^{\star})=J_{\mu}\label{eq:20}\end{equation}

where

\begin{equation}
J_{\mu}^{a}=G_{\mu\nu\,,\nu}^{a}+iq^{\star}\epsilon_{abc}V_{\mu}^{b}G_{\mu\nu}^{c}\label{eq:21}\end{equation}

is the non-Abelian $SU(2)$ generalized four-current of dyons. In
deriving Eq. (\ref{eq:20}) we have used eq. (\ref{eq:16}), (\ref{eq:18})
and (\ref{eq:19}). Thus the generalized four current of dyons in
split octonion realization leads to the abelian and non-Abelian nature
of currents in this theory.

The $\mathcal{O}$-derivative of generalized four-current 

\begin{equation}
J_{\mu}=J_{\mu}(u_{0}+u_{0}^{\star})+J_{\mu}^{a}(u_{a}+u_{a}^{\star})=\left[\begin{array}{cc}
J_{\mu}e^{0} & -J_{\mu}^{a}e^{a}\\
J_{\mu}^{a}e^{a} & J_{\mu}e^{0}\end{array}\right]\label{eq:22}\end{equation}

gives rise to

\begin{equation}
J_{\mu\Vert\mu}=0\label{eq:23}\end{equation}

which shows the resemblance with Noetherian current and describes
analogues continuity equation in abelian gauge theory.

\section{Gravito-dyons in Octonion Gauge Formalism}

Following the idea of Dowker and Roche \cite{key-19} of dual mass
playing the role of magnetic charge (Heavisidean monopole), the gravito-dyons
may also be defined as the particle carrying gravitational mass $m$
and dual mass (Heavisidean mass) $h$ having the generalized mass
$Q$ given \cite{key-28}by,

\begin{equation}
Q=m-i\, h\label{eq:24}\end{equation}

Let the space-time interval\cite{key-27},

\begin{eqnarray}
ds^{2} & = & \frac{1}{n}Trace(\eta_{\mu\nu}dx^{\mu}dx^{\nu})\label{eq:25}\end{eqnarray}

where $\eta_{\mu\nu}=\left\{ \eta_{\mu\nu}^{ab}(x)\right\} (a,b=1,2,3,......n)$is
a matrix of internal space. Then we can write $ds^{2}$as,

\begin{equation}
ds^{2}=\frac{1}{4}Trace(\eta_{\mu\nu}dx^{\mu}dx^{\nu})=\frac{1}{2}(g_{\mu\nu}dx^{\mu}dx^{\nu})\label{eq:26}\end{equation}

where $Trace$ acting upon Zorn matrix realization of split octonions.As
such, the $ds^{2}$of Eq.(\ref{eq:26}) may be considered as the line
element in space-time i.e 

\begin{equation}
\eta_{\mu\nu}=g_{\mu\nu}(u_{0}+u_{0}^{\star})+\varepsilon_{\mu\nu}^{a}u_{a}^{\star}+h_{\mu\nu}^{a}u_{a}+\left[\begin{array}{cc}
g_{\mu\nu}e_{0} & \,\,\varepsilon_{\mu\nu}^{a}e_{a}\\
-h_{\mu\nu}^{a}e_{a} & \,\, g_{\mu\nu}e_{0}\end{array}\right]\label{eq:27}\end{equation}

where $\eta_{\mu\nu}^{\dagger}=\eta_{\mu\nu}$. $\varepsilon_{\mu\nu}^{a}$and
$h_{\mu\nu}^{a}$ represents the Yang-Mill's field strengths. The
symbol $(\dagger)$ denotes the Hermitian conjugate in internal space.
The Christoffel symbol (or connection) may be written by means of
split octonion realization as,

\begin{equation}
\Gamma_{\mu\nu}^{\rho}=\Gamma_{\mu\nu}^{\rho}(u_{0}+u_{0}^{\star})+\delta_{\mu}^{\rho}(\overrightarrow{L_{\nu}}\cdot\overrightarrow{u^{\star}}+\overrightarrow{K_{\nu}}\cdot\overrightarrow{u})\label{eq:28}\end{equation}

where $(\overrightarrow{L_{\nu}}\cdot\overrightarrow{u^{\star}}+\overrightarrow{K_{\nu}}\cdot\overrightarrow{u})$
is the $\mathcal{O}$-affinity of the theory. The four indexes Ricci
tensor $R_{\rho\mu\nu}^{\sigma}$ is defined as,

\begin{equation}
R_{\rho\mu\nu}^{\sigma}=R_{\rho\mu\nu}^{\sigma}(u_{0}+u_{0}^{\star})+\delta_{\rho}^{\sigma}P_{\mu\nu}\label{eq:29}\end{equation}

where $P_{\mu\nu}$ is the $\mathcal{O}$-curvature (space-time curvature)
and is expressed as

\begin{eqnarray}
P_{\mu\nu} & = & \Im_{\mu,\nu}-\Im_{\nu,\mu}-\left[\Im_{\mu},\Im_{\nu}\right]\label{eq:30}\end{eqnarray}

which retains the usual form of Riemannian space-time geometry.

Analogous to the theory of electromagnetic (dyons), the generalized
four-potential of gravito-dyons may then be expressed as \cite{key-18},

\begin{eqnarray}
V_{\mu} & = & A_{\mu}-i\, B_{\mu}=b_{\mu}-i\, q^{\star}a_{\mu}\nonumber \\
A_{\mu} & = & b_{\mu}+h\, I\, a_{\mu}\nonumber \\
B_{\mu} & = & -m\, I\, a_{\mu}\label{eq:31}\end{eqnarray}

where $b_{\mu}$is the usual space-time gravitational four-vector
potential, $I$ is the unit matrix and $a_{\mu}$is the electromagnetic
potential. Thus the four-potential $A_{\mu}$and $B_{\mu}$ are visualized
as the four-potentials associated with the gravitational mass and
Heavisidian monopole respectively. The effective mass of gravito-dyons
is defined as

\begin{eqnarray}
M_{eff} & = & m+h\label{eq:32}\end{eqnarray}

The four-potentials $A_{\mu}$and $B_{\mu}$of Eq. (\ref{eq:31})
may then be interpreted as potential describing respectively the coupling
of Heavisidean monopole with gravitational field and that of gravitational
mass with electromagnetic field. In the absence of Heavisidean mass
$A_{\mu}$ and $B_{\mu}$ may simply be expressed as gravitational
and electromagnetic potentials respectively.

Thus we can express the generalized four-potential of gravito-dyons
by means of octonions gauge formalism given by Eq. (\ref{eq:18}).
Similarly, one can construct the generalized field tensor of gravito-dyons
in terms of split octonion as, 

\begin{eqnarray}
G_{\mu\nu} & = & g_{\mu\nu}(u_{0}+u_{0}^{\star})+\varepsilon_{\mu\nu}^{a}u^{a}+h_{\mu\nu}^{a}u^{a}=\left[\begin{array}{cc}
g_{\mu\nu}e^{0} & -\varepsilon_{\mu\nu}^{a}e^{a}\\
h_{\mu\nu}^{a}e^{a} & g_{\mu\nu}e^{0}\end{array}\right]\label{eq:33}\end{eqnarray}

where

$G_{\mu\nu}=V_{\mu,\nu}-V_{\nu,\mu}$and $G_{\mu\nu}^{a}=\partial_{\nu}V_{\mu}^{a}-\partial_{\mu}V_{\nu}^{a}+q^{\star}\epsilon_{abc}V_{\mu}^{b}V_{\nu}^{c}$having
same form of Eq. (\ref{eq:19}) for the case of non-Abelian gauge
theory of dyons associated with generalized electromagnetic fields.

Defining the generalized current of gravito-dyons as

\begin{eqnarray}
J_{\mu} & = & j_{\mu}^{(G)}-i\, k_{\mu}^{(H)}\label{eq:34}\end{eqnarray}

where $j_{\mu}^{(G)}$is the four-current associated with gravitational
mass and $k_{\mu}^{(G)}$is the four-current associated with Heavisidean
monopole. The generalized vector field $\overrightarrow{\psi_{G}}$
of gravito-dyons may be expressed as,

\begin{eqnarray}
\overrightarrow{\psi_{G}} & = & \overrightarrow{G}-i\,\overrightarrow{\mathcal{H}}\label{eq:35}\end{eqnarray}

where $\overrightarrow{G}$ is the gravitational field and $\overrightarrow{\mathcal{H}}$
is the Heavisidean field strength defined as follows in terms of two
four-potential components as, 

\begin{eqnarray}
\overrightarrow{G} & = & \frac{\partial\overrightarrow{a}}{\partial t}+\overrightarrow{\nabla}\phi_{G}+\overrightarrow{\nabla}\times\overrightarrow{b}\nonumber \\
\overrightarrow{\mathcal{H}} & = & \frac{\partial\overrightarrow{b}}{\partial t}+\overrightarrow{\nabla}\phi_{h}-\overrightarrow{\nabla}\times\overrightarrow{a}\label{eq:36}\end{eqnarray}

Thus we express the generalized field $\overrightarrow{\psi_{G}}$
in terms of generalized four-current $J_{\mu}$ as,

\begin{eqnarray}
\overrightarrow{\nabla}\cdot\overrightarrow{\psi_{G}} & = & -J_{0}\nonumber \\
\overrightarrow{\nabla}\times\overrightarrow{\psi_{G}} & = & -i\frac{\partial\overrightarrow{\psi_{G}}}{\partial t}-i\,\overrightarrow{J}\label{eq:37}\end{eqnarray}

which is the field equation (generalized Maxwell-Dirac equation) for
linear gravito-Heavisidean fields.

The $\mathcal{O}$-forms of different gauge potentials associated
with gravito-dyons may be expressed as

\begin{eqnarray}
V_{\mu}= & V_{\mu}u_{0}+V_{\mu}^{a}u^{0}+V_{\mu}u_{0}^{\star}+V_{\mu}^{a}u_{a}^{\star} & =\left[\begin{array}{cc}
V_{\mu}e^{0} & -V_{\mu}^{a}e^{a}\\
V_{\mu}^{a}e^{a} & V_{\mu}e^{0}\end{array}\right]\nonumber \\
A_{\mu}= & A_{\mu}u_{0}+A_{\mu}^{a}u^{0}+A_{\mu}u_{0}^{\star}+A_{\mu}^{a}u_{a}^{\star} & =\left[\begin{array}{cc}
A_{\mu}e^{0} & -A_{\mu}^{a}e^{a}\\
A_{\mu}^{a}e^{a} & A_{\mu}e^{0}\end{array}\right]\nonumber \\
B_{\mu}= & B_{\mu}u_{0}+B_{\mu}^{a}u^{0}+B_{\mu}u_{0}^{\star}+B_{\mu}^{a}u_{a}^{\star} & =\left[\begin{array}{cc}
B_{\mu}e^{0} & -B_{\mu}^{a}e^{a}\\
B_{\mu}^{a}e^{a} & B_{\mu}e^{0}\end{array}\right]\nonumber \\
a_{\mu}= & a_{\mu}u_{0}+a_{\mu}^{a}u^{0}+a_{\mu}u_{0}^{\star}+a_{\mu}^{a}u_{a}^{\star} & =\left[\begin{array}{cc}
a_{\mu}e^{0} & -a_{\mu}^{a}e^{a}\\
a_{\mu}^{a}e^{a} & a_{\mu}e^{0}\end{array}\right]\nonumber \\
b_{\mu}= & b_{\mu}u_{0}+b_{\mu}^{a}u^{0}+b_{\mu}u_{0}^{\star}+b_{\mu}^{a}u_{a}^{\star} & =\left[\begin{array}{cc}
b_{\mu}e^{0} & -b_{\mu}^{a}e^{a}\\
b_{\mu}^{a}e^{a} & b_{\mu}e^{0}\end{array}\right]\label{eq:38}\end{eqnarray}

As such the $\mathcal{O}$-derivative of generalized field tensor
$G_{\mu\nu}$leads to, 

\begin{equation}
G_{\mu\nu\Vert\nu}=G_{\mu\nu,\nu}+\left[\Im_{\nu},G_{\mu\nu}\right]=-J_{\mu}(u_{0}+u_{0}^{\star})-J_{\mu}^{a}(u_{a}+u_{a}^{\star})=-J_{\mu}\label{eq:39}\end{equation}

where $J_{\mu}$is the generalized split octonion current associated
with gravito-dyons similar to that of dayons in generalized electromagnetic
fields given by Eq. (\ref{eq:21}). Hence like previous case, here
the O-derivative of the generalized current vanishes and the conservation
of generalized four-current follows the continuity equation with abelian
and non-Abelian gauge structures.

\section{Conclusion}

In Yang-Mill's theory, the gravito-dyons in gravito-Heavisidean fields
plays the same role as that of dyons in electromagnetic fields. As
such, the split octonion gauge formalism demonstrates the structural
symmetry between generalized electromagnetic fields of dyons and that
of generalized gravito-Heavisidian fields of gravito-dyons. In case
of split octonions, the automorphism group is described as $G_{2}$
(an exceptional Lie group). Thus octonionic transformations are homomorphic
to the rotation group $O_{3}$. Under the $SU(3)$ subgroup of split
$G_{2}$leaving $u_{0}$ and $u_{0}^{\star}$invariant, the three
split octonions $(u_{1},u_{2},u_{3})$ transform like a isospin triplet
(quarks) and the complex conjugate octonions transform like a unitary
anti-triplet (antiquarks)\cite{key-29,key-30}.

The abelian and non-Abelian gauge structures of dyons and gravito-dyons
are discussed in terms of split octonion variables in simple, compact
and consistent way. The field equations derived here are invariant
under octonionic gauge transformations. From the foregoing analysis
one can obtain the independent theories of electromagnetism and gravitation
in the absence of each other. The justification behind the use of
octonions is to obtain the simultaneous structure of $SU(2)\times U(1)$
gauge theory of dyons and gravito-dyons in simple and compact manner.
As such, the well-known $SU(2)$ non-Abelian and $U(1)$ abelian gauge
structure of dyons and gravito-dyons are reformulated in terms of
compact gauge formulation. The $\mathcal{O}$-derivative may be considered
as the partial derivative if we do not incorporate the split octonion
variable. Split octonions are described here in terms of $U(1)\times SU(2)$
gauge group simultaneously to give rise the abelian (point like) and
non-Abelian (extended structure) of dyons. This gauge group plays
the role of $U(1)\times SU(2)$ Salam Weinberg theory of electro-weak
interaction in the absence of Heavisidean monopoles. Our enlarged
gauge group$SU(2)\times U(1)$ explains the built in duality to reproduce
abelian and non-Abelian gauge structure of dyons and those for gravito-dyons.

Acknowledgement: - O. P. S. Negi is thankful to Prof. Dr. Heinz Dehnen,
Universität Konstanz, Fachbereich Physik , Post fach M 677, D-78457
Konstanz,Germany for providing him the hospitality at Konstanz.. He
is grateful to German Academic Exchange Service(DAAD) for granting
him the fellowship under re-invitation programme.

\appendix

\section{Appendix:Octonions Variables}

An octonion is defined as,

\begin{equation}
X=X_{0}e_{0}+X_{j}e_{j},\,\,\,\,\,\,\,\,\: X_{0},X_{j}\varepsilon\mathbb{R}\label{eq:40}\end{equation}

where $(j=1,2,.....7),$$e_{j}$ are octonion unit elements satisfying
following multiplication rules \cite{key-31};

\begin{eqnarray}
e_{j}e_{k} & = & -\delta_{jk}e_{0}+f_{jkl}e_{l}\nonumber \\
e_{j}e_{0} & = & e_{0}e_{j}=e_{j}\nonumber \\
e_{0}e_{0} & = & e_{0}\label{eq:41}\end{eqnarray}

where $\delta_{jk}$ is the usual Kronecker delta symbol and $f_{jkl}$
(which was regarded as the Levi-Civita tensor for quaternions) is
fully anti symmetric tensor with

$f_{jkl}=+1\forall j,k,l=123,516,624,435,471,673,672$.

The above multiplication table directly follows that the algebra of
octonion$\mathcal{O}$is not associative i.e.\begin{eqnarray}
e_{j}(e_{k}e_{l}) & \neq & (e_{j}e_{k})e_{l}\label{eq:42}\end{eqnarray}

The commutation rules for octonion basis elements are given by,

\begin{eqnarray}
\left[e_{j},e_{k}\right] & = & -2\delta_{jk}e_{0}\label{eq:43}\end{eqnarray}

and the associator 

\begin{equation}
\left\{ e_{j},e_{k},e_{l}\right\} =(e_{j}e_{k})e_{l}-e_{j}(e_{k}e_{l})=-\delta_{jk}e_{l}+\delta_{kl}e_{j}+(\epsilon_{jkl}\epsilon_{mnp}-\epsilon_{klp}\epsilon_{jpn})e_{n}\label{eq:44}\end{equation}

Octonion conjugate is defined as,

\begin{eqnarray}
\overline{X} & = & X_{0}e_{0}-X_{j}e_{j}\label{eq:45}\end{eqnarray}

and

\begin{eqnarray}
\overline{\overline{X}} & =X; & \overline{XY}=\overline{Y}\,\overline{X}\label{eq:46}\end{eqnarray}

The norm $N$ of the octonion is defined as,

\begin{eqnarray}
N(X) & =X & \overline{X}=\overline{X}X=(X_{0}^{2}+X_{j}^{2})e_{0}\label{eq:47}\end{eqnarray}

while the inverse is defined as,

\begin{eqnarray}
X^{-1} & = & \frac{\overline{X}}{N(X)}\nonumber \\
X^{-1}X & = & XX^{-1}=1.e_{0}\label{eq:48}\end{eqnarray}

The norm given by equation (\ref{eq:47}) is non-degenerate and positively
defined (over $\mathbb{R}$) and therefore every element $X\varepsilon\mathcal{O}$
has the unique inverse element $X^{-1}\varepsilon\mathcal{O}$ .For
the split octonion algebra the following new basis is considered \cite{key-29,key-30}on
the complex field (instead of real field) i.e.

\begin{eqnarray}
u_{1} & = & \frac{1}{2}(e_{1}+ie_{4});\,\,\,\,\,\, u_{1}^{\star}=\frac{1}{2}(e_{1}-ie_{4})\nonumber \\
u_{2} & = & \frac{1}{2}(e_{2}+ie_{5}),\,\,\,\,\,\, u_{2}^{\star}=\frac{1}{2}(e_{2}-ie_{5})\nonumber \\
u_{3} & = & \frac{1}{2}(e_{3}+ie_{6});\,\,\,\,\,\, u_{3}^{\star}=\frac{1}{2}(e_{3}-ie_{6})\nonumber \\
u_{0} & = & \frac{1}{2}(e_{0}+ie_{7});\,\,\,\,\,\, u_{0}^{\star}=\frac{1}{2}(e_{0}-ie_{7})\label{eq:49}\end{eqnarray}

where $i=\sqrt{-1}$ is assumed to commute with $e_{A}(A=1,2,3,...7)$
octonion units.The split octonion basis elements satisfy the following
multiplication rules;

\begin{eqnarray}
u_{i}u_{j} & = & \epsilon_{ijk}u_{k}^{\star};\,\,\,\,\, u_{i}^{\star}u_{j}^{\star}=-\epsilon_{ijk}u_{k}(i,j,k=1,2,3)\nonumber \\
u_{i}u_{j}^{\star} & = & -\delta_{ij}u_{0};\,\,\,\,\, u_{i}u_{0}=0;\,\,\,\,\, u_{i}^{\star}u_{0}=u_{i}^{\star}\nonumber \\
u_{i}^{\star}u_{j} & = & -\delta_{ij}u_{0};\,\,\,\,\,\, u_{i}u_{0}^{\star}=u_{0};\,\,\,\,\, u_{i}^{\star}u_{0}^{\star}=0\nonumber \\
u_{0}u_{i} & = & u_{i};\,\,\,\,\, u_{0}^{\star}u_{i}=0;\,\,\,\,\,\, u_{0}u_{i}^{\star}=0\nonumber \\
u_{0}^{\star}u_{i}^{\star} & = & u_{i};u_{0}^{2}=u_{0};\,\,\,\,\, u_{0}^{\star2}=u_{0}^{\star};\,\,\,\,\, u_{0}u_{0}^{\star}=u_{0}^{\star}u_{0}=0\label{eq:50}\end{eqnarray}

Gunäydin and Gürsey \cite{key-29,key-30}pointed out that the automorphism
group of octonion is $G_{2}$ and its subgroup which leaves imaginary
octonion unit $e_{7}$ invariant (or equivalently the idempotents
$u_{0}$ and $u_{0}^{\star}$) is $SU(3)$ where the units $u_{i}$
and $u_{i}^{\star}(i=1,2,3)$transform respectively like a triplet
and anti triplet accordingly associated with colour and anti colour
triplets of $SU(3)$ group. Let us introduce a convenient realization
for the basis elements$(u_{0},u_{i},u_{0}^{\star},u_{i}^{\star})$
interms of Pauli spin matrices as

\begin{eqnarray}
u_{0} & = & \left[\begin{array}{cc}
0 & 0\\
0 & 1\end{array}\right];\,\,\,\,\,\,\,\,\,\, u_{0}^{\star}=\left[\begin{array}{cc}
1 & 0\\
0 & 0\end{array}\right]\nonumber \\
u_{i} & = & \left[\begin{array}{cc}
0 & 0\\
e_{i} & 0\end{array}\right];\,\,\,\,\,\,\,\,\,\, u_{i}=\left[\begin{array}{cc}
0 & -e_{i}\\
0 & 0\end{array}\right](i=1,2,3)\label{eq:51}\end{eqnarray}

where $1,e_{1},e_{2},e_{3}$ are quaternion units satisfying the multiplication
rule $e_{i}e_{j}=-\delta_{ij}+\epsilon_{ijk}e_{k}$.As such, for an
arbitrary split octonion $A$ we have \cite{key-11,key-29,key-30},

\begin{eqnarray}
A & =au_{0}^{\star}+bu_{0}+x_{i}u_{i}^{\star}+y_{i}u_{i}= & \left[\begin{array}{cc}
a & -\overrightarrow{x}\\
\overrightarrow{y} & b\end{array}\right]\label{eq:52}\end{eqnarray}

which is a realization via the $2\times2$ Zorn's vector matrices$\left[\begin{array}{cc}
a & \overrightarrow{x}\\
\overrightarrow{y} & b\end{array}\right]$where $a$ and $b$ are scalars and$\overrightarrow{x}$ and $\overrightarrow{y}$
are three vectors with product defined as 

\begin{eqnarray}
\left[\begin{array}{cc}
a & \overrightarrow{x}\\
\overrightarrow{y} & b\end{array}\right]\left[\begin{array}{cc}
c & \overrightarrow{u}\\
\overrightarrow{v} & d\end{array}\right] & = & \left[\begin{array}{cc}
ac+\overrightarrow{x}\cdot\overrightarrow{v} & a\overrightarrow{u}+d\overrightarrow{x}-\overrightarrow{y}\times\overrightarrow{v}\\
c\overrightarrow{y}+b\overrightarrow{u}-\overrightarrow{x}\times\overrightarrow{u} & bd+\overrightarrow{y}\cdot\overrightarrow{u}\end{array}\right]\label{eq:53}\end{eqnarray}

and $(\times)$ denotes the usual vector product, $e_{i}(i=1,2,3)$
with $e_{i}\times e_{j}=\varepsilon_{ijk}e_{k}$ and $e_{i}e_{j}=-\delta_{ij}$.
Then we can relate the split octonions to the vector matrices given
by Eq. (\ref{eq:51}). Octonion conjugation of equation (\ref{eq:52})
is defined as,

\begin{eqnarray}
\overline{A} & =bu_{0}^{\star}+au_{0}-x_{i}u_{i}^{\star}-y_{i}u_{i}= & \left[\begin{array}{cc}
b & \,\overrightarrow{x}\\
-\overrightarrow{y} & \,\, a\end{array}\right]\label{eq:54}\end{eqnarray}

The norm of an octonion $A$ is thus defined as,

\begin{eqnarray}
\overline{A} & A & =A\,\overline{A}\,=(ab+\overrightarrow{x}\cdot\overrightarrow{y})\cdot\hat{1}\label{eq:55}\end{eqnarray}

where $\hat{1}$ is the identity element of the algebra given by $\hat{1}=\hat{1}u_{0}^{\star}+\hat{1}u_{0}$.

\begin{center}\par\end{center}
\end{document}